\documentstyle[aps,prd,latexsym,amsfonts,twocolumn]{revtex}
       
\newcommand{\ket}[1]{\left|#1\right>} 
\newcommand{\bra}[1]{\left<#1\right|}

\begin{document} 
 
\title{Non-uniqueness of the $\lambda\Phi^4$-vacuum}
\author{Ralf Sch\"utzhold$^{1,*}$, Ralf Kuhn$^{1,2}$,
Michael Meyer-Hermann$^1$, G\"unter
Plunien$^1$, and Gerhard Soff\,$^1$}
\address{$^1$Institut f\"ur Theoretische Physik, Technische  Universit\"at
Dresden, D-01062  Dresden, Germany\\
$^2$Max-Planck-Institut f\"ur Physik komplexer Systeme, 
01187 Dresden, Germany\\
$^*$Electronic address : {\tt schuetz@theory.phy.tu-dresden.de}}
\date{\today}
\maketitle

\begin{abstract} 
We prove that the massless neutral
$\lambda\Phi^4$-theory does not possess a unique vacuum. Based on the
Wightman axioms the non-existence of a state which preserves
Poincar{\'e} and scale invariance is demonstrated
non-perturbatively for a non-vanishing self-interaction. 
We conclude that it is necessary to break the scale invariance
in order to define a vacuum state.
The renormalized vacuum expectation value of the
energy-momentum tensor is derived from the two-point Wightman function
employing the point-splitting technique and its relation to  
the phionic and the scalar condensate is addressed.
Possible implications to
other self-interacting field theories and to different
approaches in quantum field theory are pointed out.
\end{abstract} 
   

\section{Introduction}

Various quantum field theories supposed to describe fundamental 
interactions in
physics are scale invariant, for instance gauge field theories
describing the electromagnetic and the strong interaction. 
In many cases, however,  
one may observe a scale dependence of observables, 
i.e.~vacuum expectation values of suitable operators. 
In order to elucidate
the origin of this scale it might be important to examine the
corresponding vacuum states. 
The properties of the vacuum state of self-interacting
theories could provide a deeper understanding of the origin of inherent
scales. To start investigations in this direction it seems legitimate to
consider at first a simple but generic model scenario.
As an instructive example we focus on
the massless, neutral $\lambda\Phi^4$-theory. 

The $\lambda\Phi^4$-theory was studied extensively in the framework of
perturbation theory. 
This approach is based on the vacuum state of the
free theory, which is scale invariant in the massless case and
independent of the interaction. A modification of the
vacuum state due to symmetry breaking induced by the self-interaction 
is not attainable in perturbation theory.
The ground state of the non-interacting
Hamiltonian is unique and coincides with the free vacuum
state. However,
the interacting Hamiltonian does not necessarily possess a unique
ground state and thus an analogous identification with the exact vacuum
state of the interacting theory does not hold in general. In order to
specify this vacuum state it is essential to employ an 
appropriate non-perturbative treatment of the interaction.

During the last decades non-perturbative techniques have become
increasingly important owing to their relevance to QCD. Special attention
was devoted to the non-trivial structure of the vacuum. Especially,
there are indications for a vacuum degeneracy in the
non-Abelian SU(2)-gauge theory \cite{YangMills}. As another example
one may investigate the non-linear Liouville model \cite{liouville},
which does not possess a translationally invariant vacuum. 
 
In this paper we would like to advocate ideas along this line of reasonings.
It is our main
intention to prove a clear assertion concerning the vacuum state in
the special case of the $\lambda\Phi^4$-theory. For this purpose we 
employ the axiomatic approach of Wightman,
cf.~\cite{strocchi,wightman,streater,reed}.   

This article is organized as follows:
The Wightman axioms summarized in the appendix are utilized to deduce a
proof of the non-uniqueness of the vacuum state of 
the scale invariant $\lambda\Phi^4$-theory in Section\,\ref{Proof}. 
In
Section\,\ref{symbreak} we introduce the non-perturbative vacuum via
breaking scale invariance and evaluate the corresponding expectation
values. 
Finally we address some implications of our results.  

\section{Proof of non-uniqueness}
\label{Proof}

In this Section we provide a general proof for the non-existence of a
unique vacuum in the case of the massless and neutral
$\lambda\Phi^4$-theory. For this purpose we construct the rather
general form of the corresponding two-point Wightman
function. Utilizing the non-linear equation of motion we derive
expectation values of higher powers of the fields. If we assume that a
unique vacuum exists, these expectation values have to
vanish. In view of the postulated cyclicity of the vacuum (see
the appendix) this zero results in a contradiction for
the non-vanishing self-interaction. Consequently,
a unique vacuum cannot exist.

\subsection{Classical $\lambda\Phi^4$-theory}

The action of a massless neutral scalar field possessing a
$\lambda\Phi^4$ self-coupling is given by 
\begin{eqnarray}
{\cal A}=\int d^4x\,\left(
\frac{1}{2}\,
(\partial_\mu\Phi)
(\partial^\mu\Phi)-
\frac{\lambda}{4!}\,\Phi^4
\right)\,.
\end{eqnarray}
This theory exhibits two important symmetries. As every realistic field
theory it obeys Poincar$\rm\acute e$ invariance 
\begin{eqnarray}
x^\mu\;\rightarrow\;L^\mu_{\;\nu}\,x^\nu+a^\mu\,.
\end{eqnarray}
On the other hand the action remains unchanged under transformations
of the following form 
\begin{eqnarray}
\label{scala}
x^\mu        &\rightarrow& \Omega^{-1} x^\mu\,, \nonumber\\
\partial_\mu &\rightarrow& \Omega\,    \partial_\mu\,, \nonumber\\
\Phi         &\rightarrow& \Omega\,    \Phi\,. 
\end{eqnarray}
This scale invariance of the action is a result of the dimensionless
coupling constant $\lambda$. The latter property is also essential for the
renormalizability of the corresponding perturbation theory.
By means of Legendre transformation we derive the Hamiltonian 
\begin{eqnarray}
H=\int d^3r\left(\frac{1}{2}
\left(
\Pi^2+(\nabla\Phi)^2\right)
+\frac{\lambda}{4!}\,\Phi^4\right)\,, 
\end{eqnarray}
which is non-negative for $\lambda\geq0$. 
In situations, where the Hamiltonian is unbounded
from below -- e.g.~for $\lambda<0$ -- no ground state exists at all.  
As another example we mention the $\lambda\Phi^3$-theory where 
the occurrence of arbitrarily negative energies is already present on
the classical level.
As proven in Ref.~\cite{baym}, this instability persists for the
quantized theory.
In the case of the $\lambda\Phi^4$-theory the classical ground state
is (for $\lambda>0$) simply given by  $\Phi\equiv 0$.
Turning to the quantum prescription the situation becomes less clear.    

\subsection{The exact quantum vacuum}
\label{ExQuVa}

The main intention of this article is to show that the quantization
of the $\lambda\Phi^4$-theory described above is not unique, i.e.~it
does not maintain all the symmetries of the classical theory. 
In particular, the vacuum state and thereby the
Hilbert space constructed out of it (see the appendix) cannot be scale 
invariant. In order to prove this assertion, we {\em assume} that a unique
and hence scale invariant vacuum exists and show that this assumption
leads to a contradiction. This (fictitious) state is denoted by
$\ket{\Psi_\lambda}$ to indicate the dependence on the coupling
strength $\lambda$.    

If we assume that the vacuum would be unique, i.e.~scale and 
Poincar$\rm\acute e$ invariant, it would remain unchanged by the
unitary scale transformation $\hat S(\Omega)$, which is defined by 
(see Eq.~(\ref{scala})) 
\begin{eqnarray}
\label{scaletrafo}
\hat S(\Omega)^{-1}\,\hat\Phi(\underline x)\,\hat S(\Omega)=
\Omega\,\hat\Phi(\underline x/\Omega)\,,
\end{eqnarray}
i.e.~$\hat S(\Omega)\ket{\Psi_\lambda}=\ket{\Psi_\lambda}$. Otherwise
there exists a different vacuum, which can be derived via 
$\hat S(\Omega)\ket{\Psi_\lambda}$.
Since the scale transformation $\hat S(\Omega)$ represents a symmetry
of the classical action, the two distinct vacuum states 
$\ket{\Psi_\lambda}$ and $\hat S(\Omega)\ket{\Psi_\lambda}$ correspond
to two equivalent quantum representations of the classical theory in
this situation.   

It should be noted here that an anomalous scale dimension 
(see e.g.~\cite{schroer}) of the fields $\hat\Phi$ --  
inducing a symmetry transformation with other powers in $\Omega$ than
the one in Eq.~(\ref{scaletrafo}) -- already prevents the theory from
being scale-invariant (see also Section \ref{EOMsec}). 
But since the proof of this assertion is exactly the aim of this
Section we do not assume an anomalous scale dimension {\em a priori}. 

\subsection{Dyson argument}

As it is well known, the necessity of introducing a scale already
occurs within perturbation theory (renormalization scale). 
This observation can be interpreted as a hint for the non-uniqueness
of the quantization of the $\lambda\Phi^4$-theory.
Perturbation theory is a very powerful method that 
allows the very precise calculation of many observables within quantum
field theory for {\em small} couplings $\lambda$, e.g.~cross sections, 
etc. However, properties of the {\em exact} vacuum state for finite
values of the coupling, e.g.~$\lambda=1$, cannot be obtained rigorously
within the framework of perturbation theory. 
The main argument can be traced back to Dyson \cite{dyson} 
who applied it to QED;
we shall present in the following a modified version of the proof  
regarding the $\lambda\Phi^4$-theory.

Within perturbation theory one performs a Taylor expansion with respect to
the coupling, in our case $\lambda$.
Especially, the expansion of the exact vacuum $\ket{\Psi(\lambda)}$
would read
\begin{eqnarray}
\ket{\Psi(\lambda)}
=
\sum\limits_{n=0}^\infty
\lambda^n\,\ket{\Psi_n}\,.
\end{eqnarray}
The equal sign above is correct if and only if the infinite 
summation converges (to the exact quantity). 
But if this sum converges for some non-vanishing coupling $\lambda_0$
then it converges for all (possibly complex) values of $\lambda$
which satisfy $|\lambda|<\lambda_0$ as well. 
Accordingly, the expansion above describes an analytic function within 
the circle of convergence.
But in this situation the exact vacuum could be analytically continued to 
negative values of the coupling $\lambda$, where the Hamiltonian is
(even classically) unbounded from below.
However, in such a highly unstable scenario a 
(translationally invariant, in particular stationary) vacuum cannot exist.
This contradiction leads to the conclusion that the Taylor expansion,
i.e.~the perturbative approach, does not represent an analytic but an
asymptotic expansion. 

Therefore, perturbation theory is applicable in a sufficiently small
vicinity of the origin $\lambda=0$ -- but not for finite $\lambda$
such as $\lambda=1$.
Consequently, the fact that the scale invariance is broken in
perturbation theory does not necessarily imply that it is broken 
for the exact vacuum state corresponding to finite $\lambda$.
Instead one might imagine that the vacuum could be scale invariant at
all fixed points, lets say at $\lambda=1$ and $\lambda=0$. 
Within perturbation theory one cannot exclude this possibility -- 
instead one is led to search for non-perturbative methods.

\subsection{Wightman function\label{wight_proof}}

Poincar$\rm\acute e$ and scale invariance of the vacuum state
impose strong restrictions on the corresponding Wightman
functions. Due to the translational symmetry they may depend on
the difference of the coordinates $(\underline x-\underline x')$
only. If we 
restrict ourselves to a region away from the light-cone 
$(\underline x-\underline x')^2\neq 0$ Lorentz invariance implies,  
that merely the scalar 
$(\underline x-\underline x')^2$ enters the Wightman function. 
Taking into account the scale invariance
$W(\Omega^{-1}\underline x,\Omega^{-1}\underline x')=
\Omega^2 W(\underline x,\underline x')$
the two-point function has to adopt the following form 
\begin{eqnarray}
\label{WF1}
W(\underline x,\underline x')=
\frac{\rm const}{(\underline x-\underline x')^2}
\end{eqnarray}
for $(\underline x-\underline x')^2\neq 0$. 
By inspection we observe that the action of the d'Alembert operator
$\Box=\partial_\mu\partial^\mu$ on this function yields zero. At first
this holds away from the light cone.  

To examine additional contributions on the light cone such as
$\delta[(\underline x-\underline x')^2]$ we investigate the Fourier
transform

$\widetilde{W}$. 
Every positive $L_+^\uparrow$-invariant distribution $\widetilde\zeta$
with support in the closed forward cone 
${\rm supp}(\widetilde\zeta)\subseteq \overline{V_+}$ has to take the
form (see \cite{garding,strocchi} and \cite{reed}, Theorem IX.33)
\begin{equation} 
\widetilde \zeta(\underline k)=a\,\delta^4(\underline k)+\Theta(k_0)\,
\mu(\underline k^2)
\end{equation}
with $a \geq 0$ and a positive measure $\mu \geq 0$ with  ${\rm
supp}(\mu)\subseteq \overline{\mathbb{R}_+}$. 
In view of the Wightman axioms the Fourier transform of the Wightman
function $\widetilde W(\underline k)$ has to be represented by a special
choice of $\widetilde\zeta(\underline k)$. The above theorem allows for the
K{\"a}ll\'en-Lehmann spectral representation \cite{kallen}
of the Wightman function
\begin{eqnarray} \label{KL}
W(\underline x,\underline x')=a+\int d\mu(m^2)\,
W^{\rm free}(\underline x,\underline x',m^2)
\end{eqnarray}
where $W^{\rm free}(\underline x,\underline x',m^2)$ denotes the 
Wightman function of a free scalar field with mass $m$, 
cf.~\cite{strocchi} and \cite{reed}, Theorem IX.34. 
The imposed scale invariance of the Wightman
function $\widetilde W(\Omega^2 \underline k^2)=\widetilde W(\underline
k^2)/\Omega^2$ implies $a=0$ and
$\mu(\Omega^2\chi)=\mu(\chi)/\Omega^2$. As a consequence, if $\mu$
contributes for positive $\chi$ then it has to behave (for $\chi>0$) like
$\mu(\chi)=b/\chi$ with $b \geq 0$. However, the resulting quantity
$\widetilde \zeta (\underline k)=\Theta(k_0)\,\Theta(\underline k^2) b / 
\underline k^2$ does not
represent a well-defined distribution owing to the singularity at $\underline
k^2=0$ together with the Heaviside step-function $\Theta$. Equivalently it does
not possess a Fourier transform. This can easily be verified by
considering 
\begin{equation}
\Box\zeta(\underline x,\underline x') = -b\,{\cal F} 
\left(\Theta(k_0)\Theta(\underline
k^2)\right)=\frac{-8\pi b}{(\underline x-\underline x')^4}
\end{equation}
which yields for $(\underline x -\underline x')^2>0$. But
no scale invariant distribution exists which generates the 
r.h.s.~of the above equation
when the d'Alembert operator is applied to. On the contrary --
as we have observed in Eq.~(\ref{WF1}) -- the action of the d'Alembert
operator on the Wightman function yields zero -- at least for 
$(\underline x -\underline x')^2\neq 0$. 
As a result, the support of the measure $\mu$ can merely contain the point
$\chi=0$. There exists only one positive distribution with support at
the origin -- the Dirac $\delta$-function. Ergo, the remaining
possibility for the Fourier transform of the Wightman function is
given by     
\begin{eqnarray}
\widetilde{W}(\underline k)=\Theta(k_0)\,\delta(\underline k^2)\cdot
{\rm const}\,.
\end{eqnarray}
This quantity indeed obeys scale invariance.
In conclusion assuming a unique vacuum the d'Alembert operator 
acting on the Wightman function vanishes everywhere
\begin{eqnarray}
\underline k^2\,\widetilde{W}(\underline k)=0
\;\leftrightarrow\;
\Box\,W(\underline x,\underline x')=0\,.
\end{eqnarray}
and in particular on the light cone.

\subsection{Equation of motion}
\label{EOMsec}

The variation of the action $\delta {\cal A}=0$ leads to the non-linear
equation
\begin{eqnarray}
\label{EOM}     
\Box\hat\Phi=-\frac{\lambda}{3!}\,\hat\Phi^3\,.
\end{eqnarray}
The field $\hat\Phi(\underline x)$ is represented by an operator-valued
distribution. However, the product of two or more distributions with
the same argument, for example $[\delta(x)]^3$ is not well-defined 
in general. Consequently, the source term on the r.h.s.~of the
equation above $[\hat\Phi(\underline x)]^3$ has at first glance no
definite meaning. 
Strictly speaking, we have to define the non-linear source term
$\hat j =\Box\hat\Phi=-\lambda\hat\Phi^3/3!$ as a local
operator-valued tempered distribution.
By virtue of the equation of motion it has to obey the following
relation under rescaling, see Eq.~(\ref{scaletrafo})
\begin{eqnarray}
\hat S(\Omega)^{-1}\,\hat j(\underline x)\,\hat S(\Omega)=
\Omega^3\,\hat j(\underline x/\Omega)
\,.
\end{eqnarray}
As already discussed in Section \ref{ExQuVa}, the
occurrence of an anomalous scale dimension of the fields 
$\hat\Phi(\underline x)$ or -- more generally -- the 
introduction of a renormalization scale $\Lambda_{\rm R}$ in order to 
define $\hat j$, i.e.~$\hat j=\hat j(\Lambda_{\rm R})$,
violate this condition.
But in this case the proof of the non-uniqueness of the exact vacuum
state is already complete at this stage:
In this situation the vacuum has to depend on this
renormalization scale as well.  
Otherwise the difference of two source
terms corresponding to different scales acting on the 
vacuum (supposed to be invariant) yields zero
\begin{eqnarray}
\left(\hat j(\Lambda_{\rm R})
-\hat j(\Lambda_{\rm R}')\right)
\ket{\Psi_\lambda}=0
\end{eqnarray}  
according to the equation of motion (\ref{EOM}).
With the same arguments as used at the end of the next Section
this implies $\hat j(\Lambda_{\rm R})-\hat j(\Lambda_{\rm R}')=0$,
which contradicts the assumption of a scale dependent source.
In summary, the eventual necessity of introducing a renormalization
scale in order to define $\hat j$ would result in a dependence of
the vacuum on this scale.

\subsection{Federbush-Johnson theorem}

As shown in Sec.~\ref{wight_proof} assuming the existence of a unique 
vacuum the two-point Wightman function equals 
(up to an irrelevant pre-factor) the two-point function of the free field.
On the other hand, we may now exploit the following trivialization
theorem, which is sometimes called the Federbush-Johnson theorem:
{\em If the two-point function coincides with its free-field analogue
then the theory is free}, see
\cite{streater,federbush,jost,schroer,pohlmeyer}.  

In the following we sketch a proof of this theorem:
If the action of the d'Alembert operator on the two-point Wightman
function yields zero we may utilize the equation of motion via   
\begin{eqnarray}
\Box \Box' W(\underline x,\underline x') &=& 
\bra{\Psi_\lambda}\Box \hat\Phi(\underline x)\Box'\hat\Phi(\underline
x')\ket{\Psi_\lambda}\nonumber\\ &=&
\left(\frac{\lambda}{3!}\right)^2\,\bra{\Psi_\lambda}\hat\Phi^3(\underline
x)\hat\Phi^3(\underline x')\ket{\Psi_\lambda}\nonumber\\ &=&0\,. 
\end{eqnarray}
This equality holds for all $\underline x$ and $\underline x'$ and 
especially for
$\underline x=\underline x'$. Accordingly, we obtain
$\bra{\Psi_\lambda}[\hat\Phi^3(\underline x)]^2\ket{\Psi_\lambda}=0$, which
implies $\hat\Phi^3(\underline x)\ket{\Psi_\lambda}=0$.
The last conclusion was possible because the Hilbert space $\frak H$
possesses a positive definite scalar product, for a Fock space with
an indefinite metric additional considerations are necessary.

Now we may exploit the postulated cyclicity  
${\frak A}\ket{\Psi_\lambda}=\frak H$ of the
vacuum (see the appendix). This property implies that all states of
the Hilbert space can be approximated by polynomials of fields (smeared
with test functions) acting on the vacuum.
Utilizing analyticity arguments (theorem of identity for holomorphic
functions)  it can be shown that it is sufficient to employ test
functions with support in an arbitrary 
small open domain $\cal O$. This fact is known as Reeh-Schlieder
\cite{Reeh} theorem: 
$\overline{{\frak A}({\cal O})\ket{\Psi_\lambda}}=\frak H$. 
One consequence of this theorem is the fact that if a local
operator annihilates the vacuum, it is the zero operator, 
cf.~\cite{strocchi} and \cite{streater}.
As a result the annihilation of the vacuum 
$\hat\Phi^3(\underline x)\ket{\Psi_\lambda}=0$
again implies $\hat\Phi^3=0$, i.e.~a free theory.

In a similar way one can also show that the field does not only
satisfy the equation of motion but also the commutation relations of a
free field \cite{araki}.
This can be demonstrated via considering the quantity
\begin{eqnarray}
\hat{\mathfrak G}(\underline x,\underline x')=
\left[\hat\Phi(\underline x),\hat\Phi(\underline x')\right]-
\bra{\Psi_\lambda}
\left[\hat\Phi(\underline x),\hat\Phi(\underline x')\right]
\ket{\Psi_\lambda}
\end{eqnarray}
and an argumentation similar to the one above, see
\cite{schroer} and \cite{pohlmeyer}.   

The proof by Federbush and Johnson in Ref.~\cite{federbush} employs
canonical commutation relations and analyticity arguments but it does
not refer to the Reeh-Schlieder property, which was established later. 

A completely different argument indicating the unphysical consequences
of the annihilation of the vacuum by the source term
is based on the natural assumption that the free
theory should be recovered in the limit $\lambda\rightarrow0$. Hence,
the independence  of the identity 
$\hat\Phi^3(\underline x)\ket{\Psi_\lambda}=0$ 
of the coupling $\lambda$ is in conflict
to the fact, that $\hat\Phi^3\ket 0 = 0$ is not valid in the
non-interacting situation. 

In summary, these contradictions lead to two alternatives:
Either the self-interaction vanishes or the vacuum is not unique.
In the former case the vacuum is Poincar{\'e} and scale invariant, but
the theory is trivial, see e.g.~\cite{trivial}.    
Assuming a non-trivial self-interacting $\lambda\Phi^4$-theory
(latter case) {\em no} unique vacuum exists. 

\section{Symmetry breaking}
\label{symbreak}

As we have shown in the previous Section a regular state that obeys
all symmetries of the considered theory does not exist. 
Accordingly, the only possibility
to define a vacuum is to break at least one of the symmetries. 
Certainly one agrees that the Poincar{\'e} invariance in
fundamental field theories on a Minkowski space-time should not be
broken. Without this symmetry it is by no means 
obvious how to distinguish the vacuum from all
other states. As a consequence, we have to break the only symmetry
left, i.e.~the scale invariance. Even though the action exhibits no
specific scale, the introduced vacuum now displays a scale-dependence.
We denote the scale of the
symmetry breaking by $\Lambda_\Phi$ and the vacuum accordingly by
$\ket{\Psi_\lambda^\Lambda}$.

In the following we are going to analyze the consequences of this 
{\em Ansatz}. To investigate the relation of the vacuum state
$\ket{\Psi_\lambda^\Lambda}$  to the ground state of the theory we
have to evaluate the renormalized 
expectation value of the energy density, i.e.~the $00$-component of the
energy-momentum tensor. In conjunction with the Wightman formalism 
it is most convenient to employ the powerful point-splitting renormalization
technique \cite{wald}, which is well-established in quantum field theory
on curved space-times. With this tool we are able to calculate the
expectation value of the energy-momentum tensor and the phionic and
scalar condensates.     

\subsection{Point-splitting\label{point}}

Several interesting observables, e.g.~the energy-momentum tensor,
contain two or more fields at equal 
space-time points $\hat A(\underline x)\hat B(\underline x)$. 
Due to the singular character of the product of
two distributions with the same argument such quantities usually
diverge $\langle\hat A(\underline x)\hat B(\underline x)\rangle=\infty$. 
This necessitates an appropriate regularization and renormalization
scheme. Having at hand merely the Wightman functions as input
information the only well-known procedure, which can be
applied directly, is the point-splitting method. 

Accordingly, one at first considers the fields at distinct space-time
points $\langle\hat A(\underline x')\hat B(\underline x)\rangle<\infty$ and
takes the coincidence limit afterwards -- a method called point-splitting
regularization. In order to generate physical reasonable, i.e.~finite
(renormalized) results, those terms, which become singular in the limit
$\underline x' \rightarrow \underline x$, have to be discarded.  

The physical meaning of the renormalization scheme described above can  
be understood by considering a physical reasonable measurement process.
Realistic detectors always produce finite results. Due to the
fact, that those detectors are not point-like, but exhibit a finite
extension the corresponding expectation values are finite as well. 
A linear detector (in the free field example a one-particle detector)
can be described by the product of two fields smeared with the test
functions $F$ and $G$, see Eq.~(\ref{smear}). The response of that
detector is given by the (finite) expectation value
$\langle\hat\Phi(F)\,\hat\Phi(G)\rangle$. In order to consider
a divergent expectation value -- for instance 
$\langle\hat\Phi^2\rangle$ -- as a
limiting case of responses of suitable detectors one may proceed as
follows: At first the space-time supports of the test functions $F$
and $G$ shrink to non-coinciding points $\langle\hat\Phi(\underline
x)\hat\Phi(\underline x')\rangle$. The associated expectation value is still 
finite. Then one considers the coincidence limit 
$\underline x\rightarrow\underline
x'$, where the expectation value diverges. Accordingly, this
idealization of a physical detector exactly corresponds to the
point-splitting procedure. 

The mechanism described above can be elucidated by a simple
example. Let us consider the following {\em Ansatz} for the exact
two-point Wightman function:
\begin{eqnarray}
\label{ansatz}
W(\underline x,\underline x+\underline{\Delta x}) 
&=& \sum\limits_n a_n(\Lambda_{\rm F})\,
\underline{\Delta x}^{2n}\,\Lambda_{\rm F}^{2n+2}\nonumber\\ 
&&+\sum\limits_n b_n(\Lambda_{\rm F})\,
\underline{\Delta x}^{2n}\,\Lambda_{\rm F}^{2n+2}\,
\ln{\left|\underline{\Delta x}^2\,\Lambda_{\rm F}^2\right|}\,.\nonumber\\
\end{eqnarray} 
(For reasons of simplicity we restrict ourselves to space-like separations.)
This expansion is correct for a sufficiently well-behaving spectral
measure $\mu(m^2)$ in the K\"all\'en-Lehmann representation in
Eq.~(\ref{KL}). 
For more complicated measures further terms such as 
$\ln^2{\left|\underline{\Delta x}^2\,\Lambda_{\rm F}^2\right|}$ may
appear and create additional singularities without altering
the following considerations. Owing to the occurrence of the
logarithmic terms we had to introduce a scale
$\Lambda_{\rm F}$. As a variation of this scale induces a transfer of 
contributions from the $b_n$ to the $a_n$ terms these coefficients may
dependent explicitly on $\Lambda_{\rm F}$. This scale characterizes
the way of distinguishing the different contributions to the response
of the detector. For the moment it is completely determined by the
observer and should not be confused with the scale of symmetry
breaking $\Lambda_\Phi$, which is a property of the vacuum.  

Having at hand the explicit expression for the Wightman function we
are now able to derive the renormalized expectation value of
$\hat\Phi^2$. To this end one keeps only those terms of the above
equation, which are finite in the coincidence limit, i.e.~one obtains
\begin{eqnarray}
\langle\hat\Phi^2\rangle_{\rm ren}^{\Lambda_{\rm F}}= 
a_0(\Lambda_{\rm F})\,. 
\end{eqnarray}
Note, that the renormalized expectation values may depend on the
the scale $\Lambda_{\rm F}$ as well.   

\subsection{Operator product expansion}

In order to elucidate the physical meaning of the introduced scales
$\Lambda_{\rm F}$ and $\Lambda_\Phi$ we examine their relation to an 
important and powerful tool in quantum field theory --
the operator product expansion \cite{OPE}.
Considering the expectation value of a product of two
operators at distinct space-time points as non-local quantity it is
possible to perform an expansion into a sum of local operators with non-local
coefficients
\begin{eqnarray}
&&\left\langle
\hat A(\underline x+\underline{\Delta x}/2)
\hat B(\underline x-\underline{\Delta x}/2)
\right\rangle
=
\nonumber\\
&&\sum\limits_n C_n(\underline{\Delta x},\Lambda_{\rm F})
\left\langle\hat O_n(\underline x,\Lambda_{\rm F})\right\rangle\,.
\end{eqnarray}
Within the framework of the operator product expansion 
$\Lambda_{\rm F}$ is denoted as the
 factorization scale. 
Considering the most simple example $\hat A=\hat B=\hat\Phi$
leads us back to the Wightman function in Eq.~(\ref{ansatz}).
The operator corresponding to $(\underline{\Delta x}^2)^{n=0}=\rm const$
exactly represents the second-order scalar condensate 
$\langle\hat O_0\rangle=\langle\hat\Phi^2\rangle_{\rm ren}$.
Calculating the expectation values of the local
operators $\hat O_n$ in the vacuum $\ket{\Psi_\lambda^\Lambda}$, 
which possesses
the symmetry breaking scale $\Lambda_\Phi$, this scale obviously
enters the local quantities $\langle\hat O_n\rangle$ as well. 
By inspection we observe that
for small distances $\underline{\Delta x}^2\Lambda_\Phi^2\ll1$ the
lowest-order term is most relevant. This contribution describes
the short range behavior of the theory. On the other hand for large
distances $\underline{\Delta x}^2\Lambda_\Phi^2={\cal O}(1)$ higher-order
contributions become more and more relevant. 
Accordingly, the long range features -- mediated via the
operators $\hat O_n$ -- usually dominate in this situation. 
Evidently, the symmetry breaking scale can be envisaged as
the {\em natural scale of factorization}, which distinguishes 
between the long and short range behavior
$\Lambda_\Phi=\Lambda_{\rm F}$. 

\subsection{Observables\label{OBSERV}}

In analogy to the second-order scalar condensate 
$\langle\hat\Phi^2\rangle_{\rm ren}=a_0$ we may derive further
renormalized expectation values by means of an appropriately chosen
differential operator acting on the Wightman function in Eq.~(\ref{ansatz}).
The contributions which are finite in the coincidence limit read
\begin{eqnarray}
\label{box}
\langle\hat\Phi\Box\hat\Phi\rangle_{\rm ren}
=4\Lambda_\Phi^4(2a_1+3b_1)\,.
\end{eqnarray}
By virtue of Poincar{\'e} invariance we can deduce  
\begin{eqnarray}
\langle\partial_\mu\hat\Phi\partial_\nu\hat\Phi\rangle_{\rm ren}=
-g_{\mu\nu}\Lambda_\Phi^4(2a_1+3b_1)\,.
\end{eqnarray}
Utilizing the equation of motion (\ref{EOM}) we derive from 
Eq.~(\ref{box}) the fourth-order scalar condensate
\begin{eqnarray}
\langle\hat\Phi^4\rangle_{\rm ren}=
-\frac{4!}{\lambda}\Lambda_\Phi^4(2a_1+3b_1)\,,
\end{eqnarray}
where we have used the definition $\hat\Phi^4=\hat\Phi^3 \hat\Phi$ 
that is consistent
with the equation of motion and the Heisenberg representation. 
The factor $1/\lambda$ indicates that our non-perturbative results
cannot be obtained using elementary perturbation theory. 
The expectation value of the Lagrangian density corresponding to the
phionic condensate yields
\begin{eqnarray}
\langle\hat{\cal L}\rangle_{\rm ren}=
-\Lambda_\Phi^4(2a_1+3b_1)\,.
\end{eqnarray}
These ingredients enable us to achieve one of our main intentions --
the derivation of the energy-momentum tensor
$T_{\mu\nu}=\partial_\mu\Phi\partial_\nu\Phi-g_{\mu\nu}\,{\cal L}$. We
observe the exact cancellation of the above contributions
\begin{eqnarray}
\label{Tmunu}
\langle\hat T_{\mu\nu}\rangle_{\rm ren}=0\,.
\end{eqnarray}
Note, that for deriving this equation we merely need Poincar{\'e}
invariance (together with the point-splitting technique). 
As a counter-example one may remember the Casimir effect where
$\langle\hat T_{00}\rangle_{\rm ren}<0$ holds, even for $\lambda=0$.

The above result indeed confirms the identification of the vacuum state
$\ket{\Psi_\lambda^\Lambda}$ with a ground state. Because of 
$\hat H=\int d^3r\, \hat T_{00}$ it follows 
$\langle\hat H\rangle_{\rm ren}=0$ from the equation above. 
(The spectral condition explained in the appendix together with the
scale invariance implies $\hat H_{\rm ren}\geq 0$.) Ergo, any of the
introduced vacua $\ket{\Psi_\lambda^\Lambda}$ 
characterized by a positive value of the scale $\Lambda_\Phi$ 
may be identified as ground
states of the theory (which are therefore not unique).    

It should be emphasized, that the zero expectation value 
in Eq.~(\ref{Tmunu}) is
rather non-trivial.  
Considering a massless scalar field with a
$\lambda\Phi^n$-coupling in a $D$-dimensional space-time one arrives at
\begin{equation}
\langle T_{\mu\nu}\rangle_{\rm ren}\sim g_{\mu\nu}\Lambda_\Phi^n
\left(1-\frac{D}{2}+\frac{D}{n}\right)\,.
\end{equation}
Cancellations similar to the situation discussed above
occur exactly in those cases, 
where the theory is scale invariant, i.e.~for $2(n+D)=nD$.
In addition to the absence of any mass terms the
scale invariance implies a dimensionless coupling constant.

\section{Conclusions}

\subsection{Summary}

Utilizing the Wightman axioms we have shown for the 
scalar, massless, neutral, and
self-interacting $\lambda\Phi^4$-theory that no state exists,
which preserves Poincar\'e as well as scale invariance, i.e., all the
symmetries of the Lagrangian. Accordingly, we are lead to introduce the
non-perturbative vacuum state by breaking the scale invariance.
Consistent with the Wightman approach we employed the point-splitting
technique, which allows for an explicit evaluation
of renormalized expectation values.
Within this formalism we calculated the scalar as well as
the phionic condensate. The renormalized vacuum expectation value of
the energy-momentum tensor vanishes, which implies
that all vacua $\ket{\Psi_\lambda^\Lambda}$ are ground states.    

\subsection{Discussion}
\label{Discussion}

Raising the question about the existence of a unique vacuum state in a
field theory including a non-trivial interaction term
we focused on the real and massless $\lambda \Phi^4$-theory. 
Perturbation theory is based on the vacuum of the free theory, which
can be uniquely determined and coincides with the ground state of the 
corresponding free Hamiltonian.
It is evident from the beginning that a unique vacuum state should respect
all the symmetries of the underlying theory, i.e.~that of the Lagrangian.
Since the vacuum state 
is defined via the free theory in perturbation theory, this property of
the vacuum is established by brute force and is independent of the form of the
interaction term in the Lagrangian.

In the framework of non-perturbative
methods there is a need for the definition of a corresponding
exact vacuum state of the interacting theory. 
Naively this state should
again respect all the symmetries of the underlying Lagrangian
but now incorporating the interaction term. 
The latter will in general have some impact on the vacuum state. 
As indicated below the structure of the exact vacuum state 
becomes rather complex in comparison with the free (perturbative) 
vacuum.

As a generic example we checked, 
whether such a non-perturbative vacuum state can be found
in the $\lambda \Phi^4$-theory. To this end we started with the
definition of a vacuum state as a state, which preserves all 
the symmetries of the Lagrangian: Poincar\'{e} and scale invariance.
However, it turned out that 
the conjectured vacuum state does not allow for the
generation of all other states in the self-interacting theory by means of
field operators, and thus it is inconsistent with the property of
cyclicity of vacuum states.
Therefore the vacuum cannot be unique.
 
We conclude that the only reasonable way to define a vacuum of
$\lambda\Phi^4$-theory is to break scale invariance. 
As a consequence, the vacuum state of the theory now depends on a new scale
$\Lambda_\Phi$. We were able to interpret this scale in the
framework of operator product expansion (OPE), where the expectation
value of a field product is decomposed into a sum of products consisting of
two parts describing the long range and short range behavior,
respectively. Here the
scale $\Lambda_\Phi$ is to be identified with the factorization
scale of OPE, i.e.~with the scale separating the components of long 
and short distances.  
This is of considerable importance in theories with asymptotic
freedom for which the short distance dependent part may be calculated 
perturbatively. However, this is not the case for the 
$\lambda \Phi^4$-theory because 
of its QED-like asymptotic behavior (Landau pole).  

Finally we were investigating possible consequences of the new scale 
$\Lambda_\Phi$ and its appearance in
observable quantities of the $\lambda\Phi^4$-theory. 
Starting from a very general structure
for the two-point Wightman function, we found that the expectation
value of the energy-momentum tensor vanishes non-trivially. 
{\it Non-trivially}
means an exact cancellation of the scale dependence in both terms
contributing to the energy momentum tensor which occurs for scale
invariant Lagrangians only. This zero result confirms the notion
of the scale dependent vacuum state as a ground state of the theory.

This result may also be compared to corresponding results obtained
in the framework of perturbation theory and one may ask about
the relation of our zero result to the known trace anomalies
\cite{dilat,collins}. In Ref.~\cite{collins} the following expression
for the trace anomaly has been derived
\begin{eqnarray}
\label{Trace}
\langle T_\mu^\mu\rangle_{\rm ren}=
-\frac{\beta}{4!}\,
\langle\Phi^4\rangle_{\rm ren}\,.
\end{eqnarray}
One should be aware that within renormalization theory 
perturbative results keep the same form independently of
the momentum flow through the corresponding Feynman diagrams.
Actually, in this paper
we calculated the {\it vacuum} expectation value of the
energy-momentum tensor. Owing to the translation invariance of the
vacuum the Fourier transform of every local expectation value
contributes only at vanishing momentum. In order to compare our result
in Eq.~(\ref{Tmunu}) with Eq.~(\ref{Trace}) we have to evaluate the
quantities there -- especially the $\beta$-function -- at zero
momentum. Since the $\lambda\Phi^4$-theory obeys an infra-red fixed
point ($\beta=0$) our zero-result $\langle T_\mu^\mu\rangle_{\rm
ren}=0$ for the energy-momentum tensor is in accordance with the
calculations within perturbation theory -- even though 
$\langle\Phi^4\rangle_{\rm ren}\neq 0$. Nevertheless one should be
careful in comparing perturbative and 
non-perturbative results, as one cannot expect in general that a
non-perturbative result has a relation to any finite order
perturbative calculation. In addition the comparison of results
obtained within different renormalization procedures (i.e.~dimensional
and point-splitting) is a delicate task.  

To elucidate the properties and the complex nature of the 
non-perturbative vacuum, we may analyze this state by
considering e.g.~its content of free particles
$\hat{N}^{\rm free}_{\vec{k}}$. Applying this number operator 
to the free vacuum yields zero and it diagonalizes the free Hamiltonian
$\hat H(\lambda=0)=\hat H^{\rm free}=\int d^3r\,\hat T_{00}^{\rm free}$.
The simultaneous ground state of all these non-negative operators
$\hat{N}^{\rm free}_{\vec{k}}$ is unique and coincides with the free
vacuum. To calculate their expectation values it is sufficient to
know the two-point function. 
Owing to the deviation of the exact Wightman function of the interacting
theory from the free (scale invariant) two-point function 
(as proved in Section \ref{Proof}) at least one
expectation value differs from zero
\begin{eqnarray}
\bra{\Psi_\lambda^\Lambda}\hat N^{\rm free}_{\vec{k}}
\ket{\Psi_\lambda^\Lambda} > 0
\end{eqnarray}
indicating that the non-perturbative 
vacuum contains a non-vanishing amount of "free" scalar particles. 
This provides another hint for the non-triviality of the zero
result in Eq.~(\ref{Tmunu}). 
The non-perturbative vacuum contains exactly such an
amount of free particles that the
contributions to the energy-momentum tensor of the
interacting theory cancel.

Traditional scattering theory is based on asymptotically free particles 
in the in- and out-states. For energy ranges where
$\bra{\Psi_\lambda^\Lambda}\hat N^{\rm
free}_{\vec{k}}\ket{\Psi_\lambda^\Lambda}$ yields significant
contributions the naive application of the above formalism is not
obviously justified. Instead the propagation of the particles 
is similar to that in a medium.

The necessity of breaking the scale symmetry in a
non-perturbative approach has consequences to the application
of the path-integral formalism. The generating functional
\begin{eqnarray}
W[J]=\int {\cal D}\Phi\,\exp\left(
i\int d^4x\;{\cal L}+J\Phi
\right)\,,
\end{eqnarray}
if it exists beyond perturbation theory with the usual regular measure 
${\cal D}\Phi$, is scale invariant per
definition. So are all expectation values deduced of it.
Usually these expectation values may be identified with the vacuum
expectation values, which are then scale invariant as well. 
But this is
inconsistent with the scale dependence of the exact vacuum state.
It follows that the usual scale invariant path-integral formalism is not
naively applicable to non-perturbative analytical calculations in the case
of the $\lambda \Phi^4$-theory.
Of course, the argument presented above does not apply to lattice calculations
where the lattice spacing induces a scale which may be connected with the 
intrinsic scale of the vacuum.

In summary the results obtained so far motivate further examinations 
concerning the relation of the presented non-perturbative approach to
other formalisms. Furthermore, it might be interesting to extend
the method for the explicit non-perturbative evaluation of expectation
values -- as presented in this article -- to other observables.

\subsection{Outline}

We expect that the non-uniqueness of the vacuum state is a more
general feature, which holds true in other scale invariant
field theories as well. 
This may especially be the case for the gauge sector of
QCD -- a statement which is currently under consideration
\cite{vorbereitung}.
If so, our assertion may have consequences for the current efforts to
find a treatable approach in QCD in the medium energy
range of some ${\rm GeV}$.

The Lagrangian governing the dynamics of the gluons
${\cal L} = -G_{\mu\nu}^a G^{\mu\nu}_a /4$ is scale invariant
as the Lagrangian of the $\lambda\Phi^4$-theory. In contrast
to the latter case further difficulties arise.
The character of this field theory as a gauge field theory
implies primary and secondary constraints.
The equations of motion are more involved and contain
terms linear and quadratic in the coupling
$g$. On the other hand there is an
additional $SU(3)$-color symmetry.

In QCD the expectation value of the Lagrangian density of the gluonic
sector $\langle\hat G_{\mu\nu}^a\hat G^{\mu\nu}_a\rangle_{\rm ren}$
represents the gluonic condensate.
The calculation of this quantity in analogy to Sec.~\ref{OBSERV}
might provide some interesting insights owing to its  
considerable relevance in QCD sum rules (see e.g.~\cite{Sumrules})
and more generally for OPE.

The energy-momentum tensor of the pure gluonic sector is
traceless at the classical level. 
Then Poincar\'e invariance would imply the vanishing of its  
renormalized vacuum expectation value
\begin{eqnarray}
\langle\hat T_{\mu\nu}\rangle_{\rm ren}=0
\quad.
\end{eqnarray}
Nevertheless, in analogy to the $\lambda\Phi^4$-theory \cite{dilat,collins} 
the phenomenon of a trace anomaly occurs in QCD as well
\cite{anomaly}. Since the Yang-Mills theory possesses a low momentum
behavior, which is different from the $\lambda\Phi^4$-theory the
arguments presented in Sec.~\ref{Discussion} do not necessarily apply
in this case. This may result in a non-vanishing expectation value.

\section{Appendix: The Wightman axioms}
\label{wight}

For the free field there exist two different options to define the
vacuum, firstly as the ground state of the Hamiltonian and secondly as
the state which is Poincar{\'e} invariant. For the interacting field
the former possibility does not apply in general.   
In the following we recapitulate an axiomatic approach to quantum
field theory based on the Wightman \cite{wightman} formalism that
utilizes Poincar{\'e} invariance. 
The quantum field $\hat\Phi$ is represented by
an  operator-valued tempered distribution acting on a separable Hilbert
space $\frak H$.
The convolution of operator-valued tempered distributions with
smooth test functions of compact support yields regular operators
which generate an algebra $\frak A$. 
Poincar{\'e} transformations are mediated via unitary operators
$\hat U(\underline L,\underline a)$
\begin{eqnarray}
\hat U(\underline L,\underline a)^{-1}\,\hat\Phi(\underline x)\,
\hat U(\underline L,\underline a)=
\hat\Phi(\underline{Lx}+\underline a)\,.
\end{eqnarray}
The Hilbert space $\frak H$ possesses a cyclic 
and Poincar{\'e} invariant 
$\hat U(\underline L,\underline a)\ket{\Psi_0}=\ket{\Psi_0}$
state $\ket{\Psi_0}$
which is called the vacuum. Per definition of cyclicity, all other 
states $\ket{\Psi}$ of the
Hilbert space $\frak H$ can be created by acting an appropriate functional
$F_{\Psi}[\hat\Phi]$ on the vacuum 
\begin{eqnarray}
\ket\Psi 
&=&
F_{\Psi}[\hat\Phi]\ket{\Psi_0}\,,      
\nonumber\\
\frak H
&=&
{\frak A}\ket{\Psi_0}\,.
\end{eqnarray}
As a consequence, the expectation values of all observables in all
states can be expressed in terms of vacuum expectation values of
field operators --  the Wightman functions (reconstruction theorem). 
In order to represent a
realistic field theory the Wightman functions have to fulfill certain
properties. These axioms are presented in the following for the
example of the two-point function for a neutral scalar field $\hat\Phi$,
see e.g.~\cite{strocchi,wightman,streater,reed}.      

\subsection{Definition}

To ensure the character of the quantum field as an operator-valued
tempered distribution the two-point Wightman function  
\begin{eqnarray}
W(\underline x,\underline x')=
\bra{\Psi_0}\hat\Phi(\underline x)\hat\Phi(\underline x')\ket{\Psi_0}
\end{eqnarray}
has to be a tempered bi-distribution. 
The property of neutral fields to be described by hermitian operators
implies
\begin{eqnarray}
W^*(\underline x,\underline x')=W(\underline x',\underline x)\,.
\end{eqnarray}

\subsection{Covariance}

In order to generate a Poincar{\'e} invariant vacuum the Wightman
functions must exhibit the same feature
\begin{eqnarray}
W(\underline x,\underline x')=
W(\underline L\underline x+\underline a,\underline L\underline x'+\underline a)
\end{eqnarray}
for all translations $\underline a$ and all rotations $\underline L$ of the
restricted Lorentz group 
$L_+^\uparrow=\{\underline L:{\rm det}\underline L=1,L^0_0>0\}$,
 which contains all
transformations connected continuously to the identity, i.e.~no time
and/or space inversion. Translation invariance implies the Wightman
function to depend on the difference of the coordinates 
$\underline x-\underline x'$ only. Inside of each light cone merely 
$(\underline x-\underline x')^2$ enters the
Wightman functions due to rotational symmetry. However, they may differ
in their values inside the future and the past light cone, respectively.  

\subsection{Spectral condition}

The properties listed above allow for the Fourier transformation of
the Wightman function according to
\begin{eqnarray}
\label{SPEC1}
W(\underline x,\underline x')
&=&
{\cal F}\left(\widetilde W\right) 
\nonumber\\
&=&  
\int d^4k\,
\widetilde{W}(\underline k)\,
\exp\left(-i\underline k(\underline x-\underline x')\right)\,.\nonumber\\
\end{eqnarray}
It should be stated that all considerations employ the Minkowski
metric $g_{\mu\nu}={\rm diag}(+1,-1,-1,-1)$. To ensure the stability
of the theory the support of this Fourier transform $\widetilde{W}(\underline
k)$ has to be contained in the closed forward cone 
$\overline{V_+}=\{\underline k:\underline k^2\geq0,k_0\geq0\}$
\begin{eqnarray}
k_0<0     &\rightarrow& \widetilde{W}(\underline k)=0 \nonumber\\
\underline k^2<0 &\rightarrow& \widetilde{W}(\underline k)=0\,.  
\end{eqnarray}
$k_0$ is related to the energy and thus, the
first condition $k_0\geq 0$ prevents the system from 
collapsing. Poincar{\'e} symmetry implies the vanishing of the Fourier
transform in the whole space-like region. 

\subsection{Locality}

By means of Einstein causality space-like separated events cannot
interfere. As a result we require the fields to commute at space-like 
distances 
and therefore the Wightman functions to be symmetric in this case   
\begin{eqnarray}
(\underline x-\underline x')^2<0\;\rightarrow\;W(\underline x,\underline x')=
W(\underline x',\underline x)\,.
\end{eqnarray}
For neutral fields the Wightman functions are therefore completely real
at space-like distances.  
 
\subsection{Positivity}

Smearing the (hermitian) operator-valued distributions 
$\hat\Phi(\underline x)$ with
smooth test functions of compact support $G(\underline x)$ one acquires regular
operators   
\begin{eqnarray}
\label{smear}
\hat\Phi(G)=\int d^4x\,\hat\Phi(\underline x)\,G(\underline x)\,.
\end{eqnarray}
The absolute value squared of an operator 
$|\hat\Phi(G)|^2=[\hat\Phi(G)]^\dagger\hat\Phi(G)=
\hat\Phi(G^*)\hat\Phi(G)$ and
thereby also its expectation value are non-negative. Therefore the
Wightman functions have to obey the following positivity condition for
all test functions $G$
\begin{eqnarray}
\int d^4x\int d^4x'\,
G^*(\underline x)\,W(\underline x,\underline x')\,G(\underline x')\geq0\,.
\end{eqnarray}
Applying the Fourier transformation on this inequality the positivity 
requirement takes the very simple form in terms of the Fourier  
transform of the Wightman function
\begin{eqnarray}
\widetilde{W}(\underline k)\geq0\,.
\end{eqnarray}

\subsection{Cluster property}

The existence of a unique translationally invariant state 
(i.e.~the vacuum) $\ket{\Psi_0}$ is used (cf.~\cite{strocchi}) to
deduce the cluster property of quantum field theories
\begin{eqnarray}
\lim\limits_{\underline s^2\rightarrow-\infty}
\bra{\Psi_0}\,\hat A(\underline x+\underline s)\,
\hat B(\underline x')\,\ket{\Psi_0} 
\nonumber\\=
\bra{\Psi_0}\,\hat A(\underline x)\,\ket{\Psi_0}\, 
\bra{\Psi_0}\,\hat B(\underline x')\,\ket{\Psi_0} 
\,,
\end{eqnarray}
where $\hat A,\hat B$ are operators composed out of fields. This
property is crucial for defining the S-matrix \cite{strocchi}.   
The existence of more than one translationally invariant state
in the Hilbert space $\frak H$ would imply that
the cluster property does not hold in general. 
However, for operators associated with physically meaningful events
the cluster property should remain valid, because events at large
space-like distances are asymptotically uncorrelated.
Recalling the scale-dependence and thereby non-uniqueness of
the vacuum of the considered $\lambda\Phi^4$-theory one is lead to the
question whether the cluster property is satisfied in this case.
Since the Hilbert space is constructed starting from the cyclic vacuum
${\frak H}(\lambda,\Lambda_\Phi)={\frak A}\ket{\Psi_\lambda^\Lambda}$
it may also depend on the scale. 
The remaining question is, whether different vacuum states 
corresponding to different scales belong to the same Hilbert space or
not, i.e.~whether 
\begin{eqnarray}
\ket{\Psi_\lambda^\Lambda}
\in
{\frak H}(\lambda,\Lambda_\Phi')\,.
\end{eqnarray}
Indeed, it is possible that different values of the scale $\Lambda_\Phi$
correspond to distinct Hilbert spaces ${\frak H}(\lambda,\Lambda_\Phi')$, 
which are not connected by local excitations. Such a situation, where
different global features generate distinct Hilbert space representations
(super-selection sectors), occurs for example in field theories at
different values of the temperature.
If the physically realized vacuum state coincides with such a vacuum
state -- which generates a Hilbert space containing only one
translationally invariant state -- then the cluster property still holds.  

\bigskip

{\bf Acknowledgment}

\bigskip

The authors are grateful to H.~Araki, D.~Diakonov, R.~Jackiw,
F.~Krauss and 
D.~J.~Schwarz for valuable conversations and kind comments.
R.~S., R.~K., and G.~P.~acknowledge fruitful discussions with K.~Sailer
and Z.~Schram during their stay at the Department of Theoretical
Physics at the University of Debrecen, Hungary. 
This visit was supported by M{\"O}B and DAAD. 
Financial support from BMBF, DFG and GSI is gratefully acknowledged. 

\addcontentsline{toc}{section}{References}


\begin{thebibliography}{99}
\bibitem{vorbereitung} These considerations will be established in
detail in a forthcoming publication.
\end{thebibliography}
\end{document}